# TCO Nanostructures as building blocks for nanophotonic devices in the infrared

Shi Qiang Li*[a], Leonidas E. Ocola[b], Robert P. H. Chang[a]

[a]Department of Materials Science and Engineering, Northwestern University, 2220 Campus Drive, Evanston, IL, USA, 60208
[b]Center for Nanoscale Materials, Argonne National Laboratory, Argonne, IL, USA, 60439

## ABSTRACT

Transparent conducting oxides (TCOs), in general, are degenerated semiconductors with large electronic band-gap. They have been widely used for display screens, optoelectronic, photonic, and photovoltaic devices due to their unique dual transparent and conductive properties. In this study, we report in detail a technique that we developed to fabricate single crystal TCO nanorod arrays with controlled conductivity, height, and lattice spacing in a simple one-zone tube furnace system. We demonstrate how novel photonic/plasmonic properties can be obtained by selecting unique combinations of these basic parameters of the nano-rod arrays.

## 1. INTRODUCTION

In last decade, tremendous excitement has been generated in research with plasmonic materials and metamaterials, which shows great potential for a wide array of applications, such as sensing, switching, communication, and etc.[1-4] However, most of the research efforts on nanophotonics, especially plasmonics, were devoted to metallic materials such as silver, gold, and copper, owing to the resonant wavelength lying at visible to near infrared regime. Recently, the quests of searching for other materials have been started[5] and a lot of excitement has been generated as these materials can be complimentary to metals, functioning at other wavelength regime, such as mid- to far- infrared.[6-11]

Among these materials, transparent conducting oxides are a class of materials which possess some unique characteristics that metals don't offer; they are stable, with tunable plasma frequencies, and often free of other loss mechanisms at the infrared regime. Furthermore, for nanophotonics applications, it is desirable to fabricate structures with controlled size, periodicity, crystallinity, geometry, and alignment. Consequently, it is important to develop nanofabrication techniques to achieve these.

For our study, we have systematically studied how to control the above mentioned properties on fabricating indium-tin-oxide (ITO) nanowires in a typical one-zone tube furnace that is available in many research labs. Have mapped out the influence of the fabrication parameters, in particular, oxygen flow rate and temperature, we proceeded to correlate the optical measurement results with their structures, subsequently demonstrated the potential of using TCO nanostructures as building blocks for nanophotonics in infrared. The techniques developed in this study are also transferrable to systems other than ITO with modifications.

## 2. EXPERIMENTAL

### 2.1 Fabrication setup

ITO nanowire growth has been studied previously by several groups.[12-16] Of particular interest, Wan *et al*[16] has demonstrated growing well aligned nanowires on substrates with excellent lattice match and was repeated in our previous study.[9] In this work, we study in-depth the influence of various parameters on the morphology of the formed nanostructures on both lattice matched substrate (Yittria Stablized Zirconia (YSZ)) and substrate without lattice match (silicon). A home-made quartz tube furnace was used in this study, which was designed with fast quenching capability (3 seconds from 900°C to 500°C) to minimize unintentionally formed product during cooling. Mixture of indium and stannous oxide at atomic ration 9 to 1 was used as source of indium and tin elements. Source temperature was kept at 900°C. The other gaseous precursor used is oxygen gas diluted with nitrogen. As a one-zone furnace is used in this study, there is a temperature gradient at the end of the furnace (EOF) (see Figure 1). This temperature gradient has been utilized to learn how the product-formation is influenced by temperature by putting the substrates at different regions downstream of the furnace to collect the products formed. The temperature profile was pre-determined and plotted as the inset in Figure1. For the convenience of description, we define epitaxially grown ITO on YSZ (100) substrate as Type A, while randomly grown ITO nanowires on silicon as Type B, as shown in Figure 1. To grow type A ITO nanowires, a very thin film of gold (1 nm) was used as catalyst.

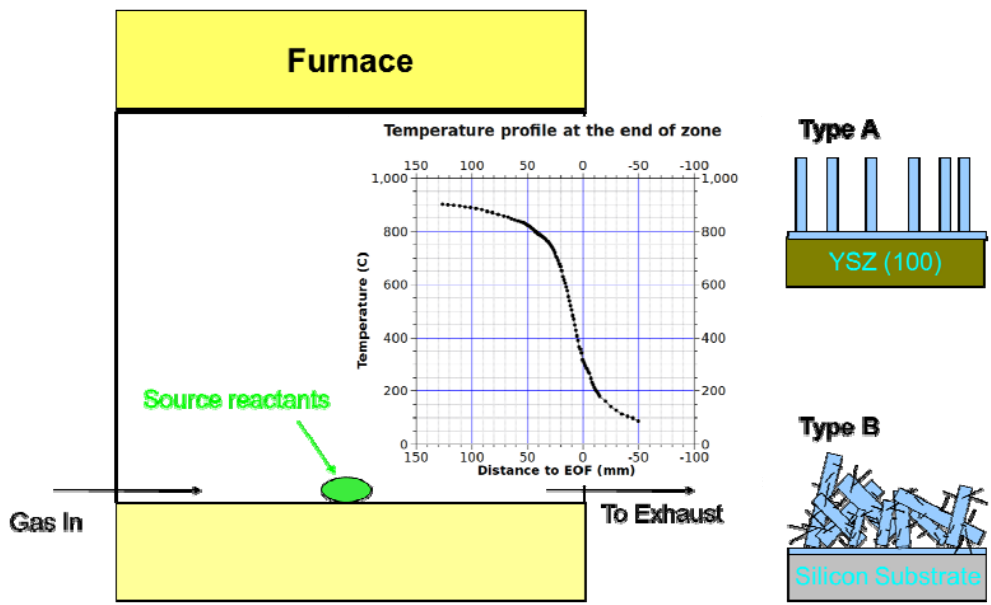

Figure 1. Schematic Diagram shows the setup of furance. Width of the furance was not drawn to scale in order to show the temperature profile at the end of furnace. NWs grown on silicon substrate are generally branched and randomly distributed. On the other hand, NWs grown on YSZ (100) are vertically aligned. EOF denotes the end of the furnace.

Optical measurements with polarized light were performed on an FTIR microscope. The polarization was achieved through a polarizer and controlling the asymmetric beam path in the cassergrain objective, which was described in details in our previous publication.[9] Scanning Electron Microscopy (SEM) was done with Hitach 4800 and Transmission Electron Microscopy (TEM) was performed on Joel 2100F FETEM with GIF camera. Patterning of the gold catalyst was done with Jeol 9300X. Gold films were thermally evaporated in Lesker Nano Thermal Evaporator.

## 3. RESULTS

In a typical one-zone quartz-tube furnace, formation of nanomaterials at different regions of the tube is known and studied by various groups.[17-19] In our system, the region downstream from the solid precursors is analyzed. It was found out that if no heterogeneous catalyst is used, product formation only happens at two regions. Firstly, green or brownish product forms at the immediate downstream of the solid precursors, which appears as a spill-like pattern, gradually decrease in the amount of product formation with the distance away from the sources. And no visible product can be seen after 50 mm away from the precursors. At this zone, the metal evaporates in a rapid rate and the ambient atmosphere is full of supersaturated indium, stannous oxide and tin vapor, which in contact with oxygen gas flow-in, leading to the formation of the product. SEM results show ITO whiskers and ITO films form at this region, depending on the substrate used.

The second region of product formation locates at the end of the furnace, when the temperature of the quartz tube is below 750°C (this critical temperature also varies somewhat with oxygen flow rate).

## 3.1 The formation of random nanowires (Type B)

The second region of product formation is our interest in this study. By putting blank silicon substrates at different positions at EOF, the morphology variation with respect to temperature change is mapped out and they are shown in Figure 2. Nanowires show increasing branching and decrease of diameter with the decrement of temperature. From the kinetics point of view, this can be explained; the nucleation rate of nanowires increases and critical nuclei size decreases exponentionally to the power of $-1/kT$.[20] Furthermore, they tend to form heterogeneously on surfaces as surface energy provides further driving force for nucleation. Small nanowires are easier to nucleate at low temperature and are prone to form on surfaces, especially homo-epitaxially on ITO nanowires themselves. However, lower temperature also reduces the thermal energy provided to overcome the energy barrier for oxide formation and leads to lower diffusion rate. Indeed, not oxidized Indium metal was observed forming at the inner wall of tube outside the furnace and low crystallinity nanowires are found at lower temperature region.

In a word, by controlling the oxygen follow rate and temperature of furnace, it is possible to control the diameter of the nanowire formation from 10 nm (500°C, 2.8 sccm $O_2$) to 1 micron (750°C 0.2 sccm $O_2$) and control their branchings, as shown in Figure 2.

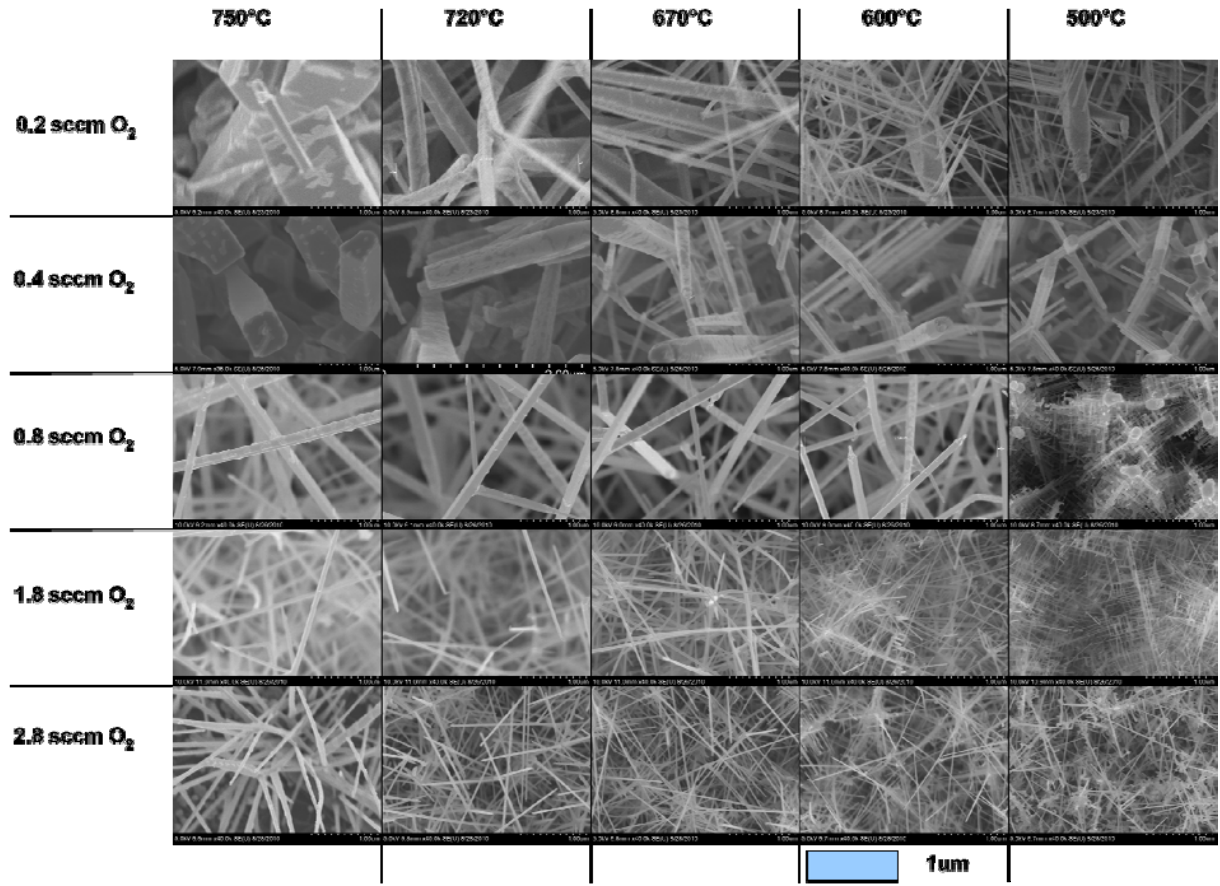

Figure 2. Matrix of SEM micrographs of TYPE B ITO NWs formed at different tempreature and oxygen flow rate on silicon substrates.

At high temperature ( > 600°C) and low oxygen ( < 1 sccm) case shown in Figure 2, it can be observed that there are particulate attachments to the nanowires. In order to find out what it is and also the crystallinity of the branched nanowires, TEM was performed. Besides that, a typical branched nanowire was also picked for the TEM study. They are shown in Figure 3. From high resolution TEM images and selected area electron diffraction (SAED) patterns at the junction of the branching shown in the inset of Figure 3(a), it is observed that the nanowires have good crystallinity and the branching wires are grown epitaxially with the main wires. The nanowires are always growing along [001] direction in accordance to the bixbyite crystal phase of indium oxide with lattice constant of 1.01nm.

To study the particulate formation on the nanowires, the sample formed with 0.8 sccm $O_2$ and located at 750°C was used for TEM due to the relatively small diameter, which will provide better electron transmission. The particulate and nanowire interface was investigated at high resolution TEM (Figure 3(b)). When beam was aligned to ITO [001] zone axis, electron diffraction pattern was collected from the interface. Two sets of diffraction patterns are identified.

The square patterns are easily told to be from ITO nanowire by comparing it with Figure 3(a). On the other hand, the other set of pattern consists of two-fold streak-shape pattern. EDS results show that at the particulate position, tin element shows a much higher concentration than indium. Possible tin dioxide crystal structures were looked up in International Crystal Structure Database (ICSD) database, and it was found that this two-fold pattern matches with rutile-phase tin-dioxide diffraction pattern. By indexing the diffraction pattern, it can be seen that the tin dioxide (020) plane is aligned with ITO (020) plane. Having retrieved the bulk lattice constant data from ICSD card number 50848 (for ITO) and 9163 (for rutile $SnO2$) , the lattice structures of ITO and tin dioxide were compared and it tells that (020) plane of tin dioxide is 2.35Å. It is less than one fourth of the lattice constant of ITO. Therefore, we suspect that, at the interface, the lattice of particulate is distorted to match the nanowire lattice and the distortion gradually relaxes away from the interface. This distortion causes the observed streak diffraction pattern.

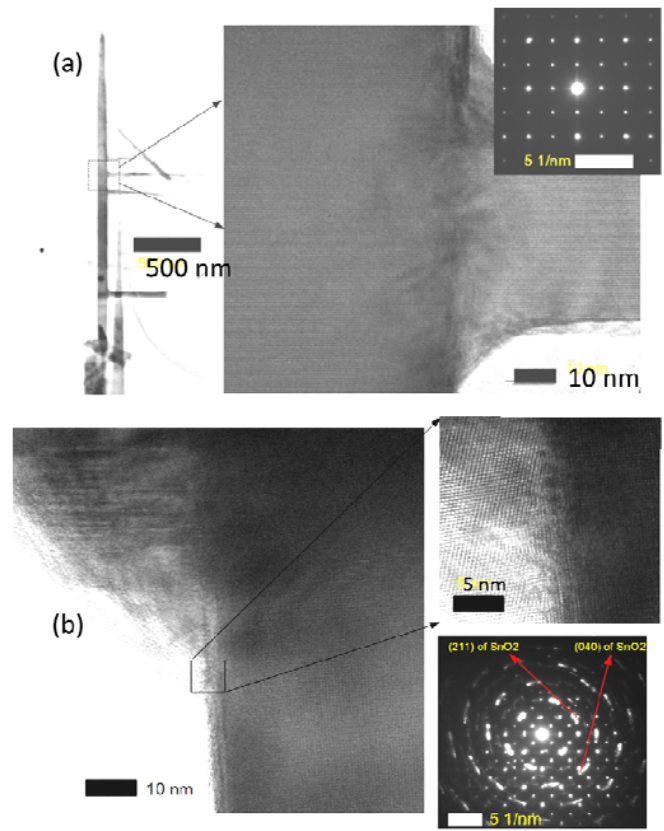

Figure 3. TEM micgraphs and diffraction patterns of branched nanowires (a) and nanowires with particulate attachments (b).

## 3.2 Vertically aligned nanowires grown on YSZ (Type A) and their optical properties.

In our furnace setup, we have identified that the ideal region for vertical aligned ITO nanowire growth with size and position control to be the region of the tube which is in between the two regions of product formation without heterogeneous catalysts. The positions where nanowires seed on the substrate in this region are defined by the position of the catalyst, because nucleation without catalyst is kinetically hindered at that temperature regime. The formation of aligned nanowire array was achieved on (100) YSZ substrate. (100) YSZ substrate was chosen as it has almost perfect match with the lattice parameter of ITO.[16] By patterning the gold catalyst, aligned nanowire array can be grown following the patterning. A resulting vertically aligned nanowire array from patterned catalyst is shown in Figure 4. As the optical studies of these periodically patterned nanorod arrays were reported by Li and *et al* elsewhere,[9] it will not be pursued further here. Rather, we focus on nanowire growth without patterning, which can be fabricated easily as the catalyst used is only thermally evaporated gold film.

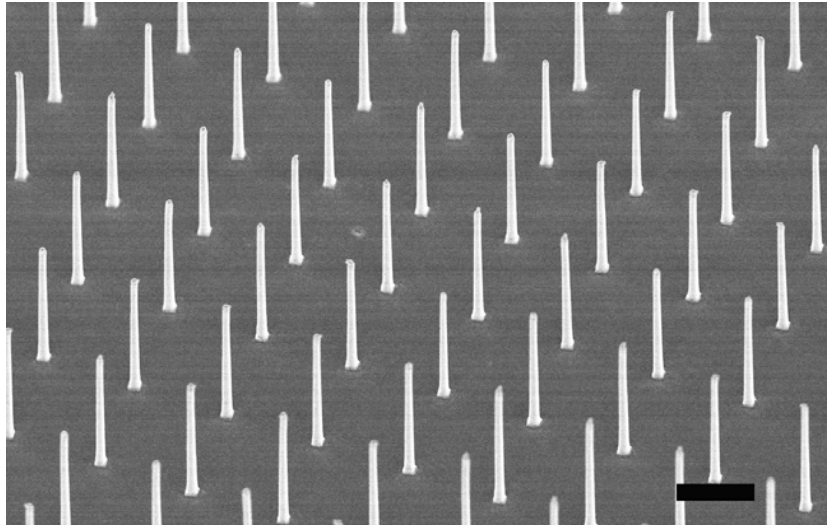

Figure 4. SEM image of a typical ITO nanorod array: This image was taken with 30° tilting to reveal the third dimension of the vertical aligned nanorods. The nanorods are 150 nm thick, 6 μm long. The lattice constant of this array is 3 μm and the scale bar is 2 μm.

By tweaking with the growth time and oxygen flow rate, a series of vertically aligned nanowires formed, with 1nm gold film as catalyst on YSZ substrates. The SEM images of the grown nanowires are shown in Figure 5 on the right hand side. Together with the aligned nanowires, three other SEM images were shown in (a), (b), and (i). For the leaves on ITO film shown in Figure 5(b), it was formed at 1 sccm $O_2$ flow rate at 40 mm downstream of the solid precursors without catalyst. The film shown in Figure 5(a) was formed under the same condition but at 50 mm downstream of the solid precursors.

Due to the alignment of the nanorods, it is expected that the interaction of aligned nanorod array with light will be sensitive to the polarization of the light.[9] Reflectance spectra with polarized light were measured with an FT-IR microscope with 37° incident angle with respect to the normal of the film.(The setup of the microscope was reported elsewhere)[9] It was found that the ITO film shows expected reflectance at the infrared region due to the free-carrier behavior. Finite-Difference-Time-Domain (FDTD) simulation[21] was performed on the sample, and the resulting spectrum was plotted as green curve in Figure 5(a), from which the plasma frequency was derived to be about 2 eV. No significant polarization dependence was observed for the film.

The *p*-pol reflectance curve shows a small dip at around 3300 nm for leaves on ITO film shown in Figure 5(b). This was not observed for *s*-pol spectrum, which is an indication of a polarization dependent resonance.

When the measured samples contain vertically aligned nanowires shown in Figure 5(c) to (h), more features are observed. Firstly, the reflectance drops significantly, at wavelength between 1667 nm to 2000 nm, independent of polarization, which is an indication of localized surface plasmon resonance (LSPR) along the short axis of the nanowires, which can be excited by both polarizations at this incident angle. This polarization independent feature extends to even longer wavelength as the nanowires become taller from Figure (c) to (h), which is due to the near field interaction of the nanorods get stronger with increased nanowire height and density.[9] For *p*-pol reflectance of these nanowire arrays, there are also dips similar to what was observed for leaves on ITO film. With the increase of wire length, features get stronger and multiple dips start to form. As these dips only exist for *p*-pol reflectance, it is related to the component of the electric field normal the surface of the film. They are attributed to the LSPR along the long axis of the nanowires. As nanowires get even longer, multipolar modes start to show, subsequently leading to more dips seen in Figure 5(f-h). For randomly aligned nanowires shown in Figure 5(i), no polarization dependent feature was observed, as expected.

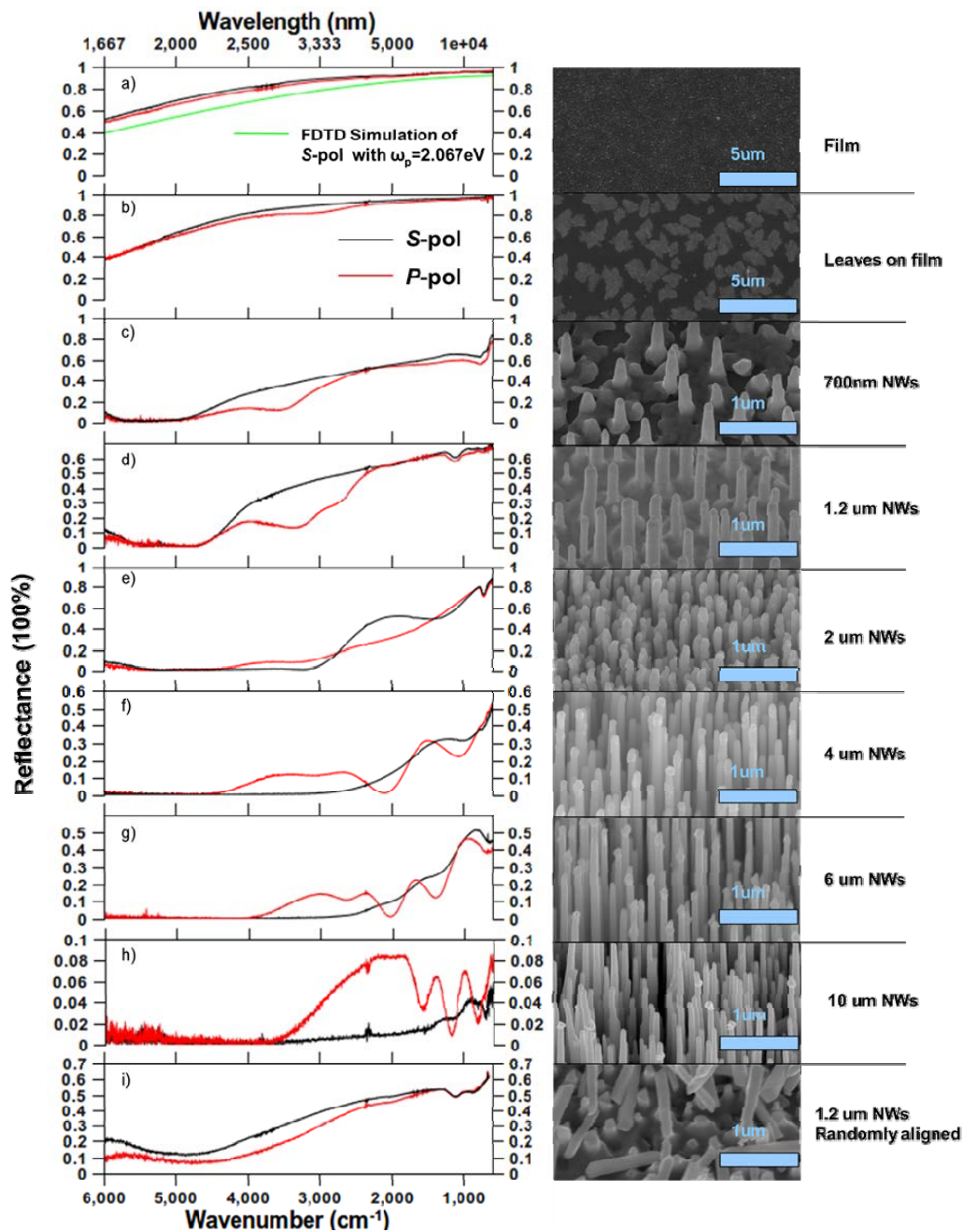

Figure 5. a to h are FTIR spectra of s- and p- incident light using setup show in Figure 6 for aligned NWs of different heights. The green curve in Figure a) is the FDTD-simulated reflectance curve for ITO thin film using Drude model with plasma frequency 2.067eV and 100nm thickness. On the right side are the corresponding microstructures viewing with 30 degrees tilting. In plot i), spectra are from the same sample in Figure d) except that the alignment is destroyed.

## 4. CONCLUSIONS

In this paper, we described a nanofabrication technique developed to produce ITO nanostructures with integrated bottom-up and top-down approaches, which can be simply implemented in any one-zone quartz tube furnaces. The nanostructures synthesized are single-crystalline, and with controlled diameter and morphology by fine-tuning the critical parameters (oxygen flow rate and temperature). We have also identified the parameters to fabricate nanowires as aligned uniform arrays arranged into different lattices and geometries or aligned but randomly placed. It was found that the alignment of the nanorods have given interesting polarization dependent optical properties, which can be connected to the localized surface plasmon resonances on anisotropic plasmonic structures. Our work provides a preliminary framework for fabricating TCO nanostructures and the methods can be transferred to other material systems. The interesting optical properties observed show great potential for designing novel nanophotonic devices working in the infrared region.

## 5. ACKNOWLEDGEMENT

We are grateful to the NUANCE facility center at Northwestern University for FTIR microscopy, SEM, TEM, and EDS measurements. The NUANCE Center is supported by NSF-NSEC, NSF-MRSEC, Keck Foundation, the State of Illinois, and Northwestern University. We also thank Professor Lincoln Lauhon, Dr Xinqi Chen, Dr. Shuyou Li, Dr Jingsong Wu and Ben Mayer for helps and discussions on characterizations. E-beam lithography was performed with Jeol-9300 in Center for Nanoscale materials in Argonne National Laboratory. Use of the Center for Nanoscale Materials was supported by the U. S. Department of Energy, Office of Science, Office of Basic Energy Sciences, under Contract No. DE-AC02-06CH11357.